\documentstyle[aps,twocolumn,epsf,rotate]{revtex}
\begin{document}
\draft
\twocolumn[\hsize\textwidth\columnwidth\hsize\csname
@twocolumnfalse\endcsname

\title{A numerical and analytical study\\ 
of two holes doped into the 2D t--J model.}
\author{A. L. Chernyshev\cite{perm}$^{\dag}$, P. W. Leung$^{\dag\dag}$, 
and R. J. Gooding$^{\dag}$} 
\address{$^{\dag}$ Dept. of Physics, Queen's University, Kingston, Ontario, 
Canada K7L 3N6}
\address{$^{\dag\dag}$ Physics Dept., Hong Kong University of Science 
and Technology, Clear Water Bay, Hong Kong}
\date{\today}
\maketitle
\begin{abstract}
Exact diagonalization numerical results are presented for a 
32-site square cluster, 
with two holes propagating in an antiferromagnetic background described by the 
$t$-$J$ model. We characterize the wave function of the lowest energy bound state found 
in this calculation, which has $d_{x^2-y^2}$ symmetry.
Analytical work is presented,
based on a Lang-Firsov-type canonical transformation derived quasiparticle
Hamiltonian, that accurately agrees with numerically determined values
for the electron momentum distribution function and the pair correlation 
function.  We interpret this agreement as strong support for the validity of 
this description of the hole quasiparticles.
\end{abstract}
\pacs
{PACS: 71.27.+a, 
75.10.Jm, 
75.40.Mg  
}
]
\narrowtext

The behaviour of mobile holes in an antiferromagnetic (AF) spin background
has been a subject of intensive study, in part because of its 
possible connection to high temperature superconductivity \cite{pwa87}.
The simplest microscopic representation of this problem 
is the so-called $t$-$J$ model \cite {elbio94,pwa97},
which captures the important AF correlations of an ``AF metal'' \cite{pwa97}.
At present only the one hole problem is well understood. That is, 
results obtained from a variety of analytical techniques agree with
one another and lead to the spin-polaron concept for the $t$-$J$
model charge carriers. 
In particular, analytical work based on the self-consistent
Born approximation (SCBA) \cite {scba} is found to be in excellent
agreement with unbiased, exact diagonalization (ED) numerics
\cite{lg95} on a large, cluster with full square symmetry. 

In this report we focus on two holes in the same $t$-$J$ model.
This problem is important because (i) it allows for the possibility of
two-particle bound states, and (ii) because it is the simplest problem
that allows one to study the interactions between charge carriers
in an AF background. At present minimal analytical work has been
done on this problem, and only small clusters have been treated numerically.
So, a complete understanding of this problem is far from at hand.

The aim of our report is twofold. 
Firstly, we combine analytical and numerical results of
the two-hole problem to provide a comprehensive study of this 
important system. We show, for the first time, the excellent agreement
between the results of these approaches.
Secondly, we argue that the spin-polaron language provides a natural
description of the physics of the system and a consistent way of 
interpreting of the numerical results. 

Computationally we have managed, for the first
time, to determine the two-hole ground state for two holes doped
into the 32-site square cluster used in our previous numerical
work \cite {lg95}. In the zero crystal momentum subspace
we obtain only one bound state, and it
has $d_{x^2-y^2}$ symmetry. 
We have characterized the ground state by
evaluating a number of important expectation values, notably
the electron momentum distribution function, and the (spatial)
pair correlation function.

We have found that the two-hole ground state derived from an effective 
quasiparticle Hamiltonian originally proposed in Ref. \cite {bel97} may be 
used to calculate the same expectation values that were obtained numerically 
via ED. Further, these quantities are remarkably similar to those obtained 
via ED, giving strong support to the appropriateness of this quasiparticle 
Hamiltonian. From the analytical point of view the low-energy physics of 
the two-hole system is generally described as a system of moderately
interacting spin polarons. This analytical work shows that the dominant 
effective interactions between spin polarons come from the short-range
nearest-neighbour static attraction and spin-wave exchange.
The purpose of this paper is to present our new ED results,
and to subsequently provide support for this description of the 
internal structure of the quasiparticles and their interactions.

We consider the $t$-$J$ model with Hamiltonian
\begin{equation}
{\cal H} = -t\sum_{\langle ij\rangle\sigma}(\tilde{c}^\dagger_{i\sigma}
\tilde{c}_{j\sigma}+{\rm H.c.})+J\sum_{\langle ij\rangle} 
({\bf S}_i \cdot {\bf S}_j
-\frac{n_in_j}{4}),
\label{eq:hamiltonian}
\end{equation}
where $\langle ij\rangle$ denotes nearest neighbour sites,
$\tilde{c}^\dagger_{i\sigma}$, $\tilde{c}_{i\sigma}$ are the constrained
operators.
We dope a 32-site square cluster with
periodic boundary conditions with two holes away from half filling. 
We focus on a realistic value of the AF exchange $J/t = 0.3$,
although many other ratios have been analyzed.
The eigenstates with zero total crystal momentum are classified 
according to their symmetry under the square point group $C_{4v}$.
We find that the $d_{x^2-y^2}$ state has the lowest energy,
which is consistent with calculations using smaller
lattices \cite{prd94}.
However, the binding is rather weak: for the ratio of
$J/t = 0.3$ we find that $E_b/t = -0.05$.
For our cluster, the binding energy becomes positive in the range
$0.2 < J/t <0.3$.

\begin{figure}
\unitlength1cm
\epsfxsize=6.5cm
\begin{picture}(8,7)
\put(-0.3,0.7){\rotate[r]{\epsffile{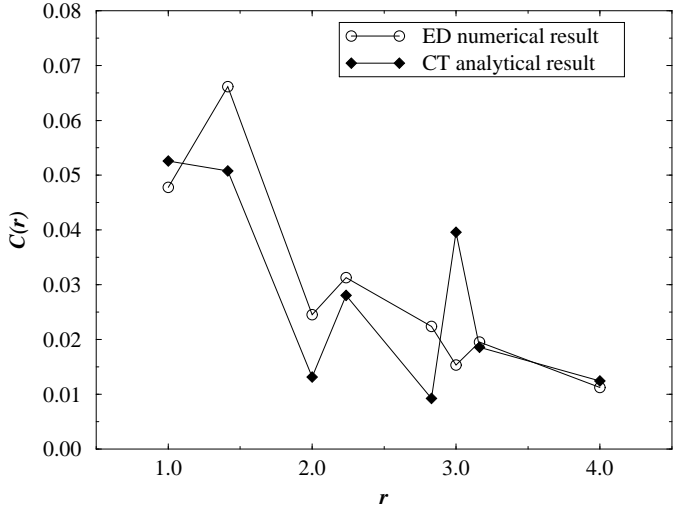}}}
\end{picture}
\caption{The spatial correlation function, $C(r)$, for two holes doped
into a square lattice described by the $t$-$J$ model, for $J/t = 0.3$.
Our ED results and the analytical results (mapped back onto a 32-site 
square lattice) are shown. The solid lines are simply a guide to the eye.}
\label{fig:Cofr}
\end{figure}

The spatial distribution of holes may be characterized by
the pair correlation function, defined as 
\begin{equation}
C(r)=\frac{1}{N_hN_E(r)}\sum_{i,j}
\langle(1-n_i)(1-n_j)\delta_{|i-j|,r}\rangle,
\label{eq:cofr}
\end{equation}
where $N_h$ is the number of holes, and $N_E(r)$
is the number of equivalent sites at a distance $r$ from any given site.
Results are shown in Fig. \ref{fig:Cofr}.

The electron momentum distribution has also been calculated:
$\langle n_{\bf q} \rangle \equiv
\langle n_{{\bf q}\uparrow} \rangle =
\langle n_{{\bf q}\downarrow} \rangle =
\langle \tilde{c}^\dagger_{{\bf q}\sigma}
\tilde{c}_{{\bf q}\sigma}\rangle$, where $\tilde{c}^\dagger_{{\bf
q}\sigma}$, $\tilde{c}_{{\bf q}\sigma}$
are the Fourier transform of the constrained operators,
and our results are shown in Table I.
This quantity has been studied in detail in smaller systems
\cite{smaller_size_nq}; however, our 32-site lattice results are important 
for such a study because the important wave vectors along the 
AF Brillouin zone (ABZ) edge [from $(0,\pi)$ to $(\pi,0)$]
are present in our cluster.  Notably, in this ground state
we do not find
any evidence of hole pockets at the single-hole
ground state momentum $(\pi/2,\pi/2)$. In fact, along the ABZ edge,
$\langle n_{\bf q} \rangle$ has a {\em maximum} at $(\pi/2,\pi/2)$.

The same problem can also be approached analytically. 
Theoretical studies of the $t$-$J$ model have resulted in a
clear understanding of the nature of the 
low-energy excitations for the system near half filling \cite {elbio94}.  
The charge carrier created by a hole introduced in an
AF background is described as a spin polaron, {\em viz.} 
as a quasiparticle consisting of the hole and a cloud of spin excitations. 
The wave function of the spin polaron in an AF background
can be written as
\begin{eqnarray}
\label{eq:spwf}
 |{\bf{k}}\rangle = \tilde{h}_{\bf k}^{\dag}|0\rangle 
=\biggl[a_{\bf k}h_{\bf k}^{\dag} + \sum_{\bf q}b_{\bf 
k, q} h_{\bf k-q}^{\dag}\alpha_{\bf q}^{\dag}
+ \dots \biggr] 
|0\rangle \ ,
\end{eqnarray}
in terms of the AF magnon and spinless hole creation operators,
$\alpha_{\bf q}^{\dag}$ and $h_{\bf k}^{\dag}$, respectively \cite {george}. 

Quite recently a new approach to the two-hole problem in the $t$-$J$ model
has been developed \cite{bel97}. It used a generalization of the canonical 
transformation approach of the Lang-Firsov type. An effective Hamiltonian
for the spin polarons, which includes both types of 
hole-hole interactions, has been obtained and its general form 
is given by
\begin{eqnarray}
\label{eq:belham}
{\cal H}_{eff}=&&\sum_{\bf k}E_{\bf k}
\tilde{h}_{\bf k}^{\dag} \tilde{h}_{\bf k}+ \sum_{\bf q} \omega_{\bf q}
\alpha_{\bf q}^{\dag} \alpha_{\bf q}\\
&&+ \sum_{{\bf k},{\bf k^{\prime}},{\bf q}}
V_{{\bf k},{\bf k^{\prime}},{\bf q}}
\tilde{h}_{\bf k-q}^{\dag} 
\tilde{h}_{\bf k^{\prime}+q}^{\dag} \tilde{h}_{\bf k^{\prime}} 
\tilde{h}_{\bf k}\nonumber\\
&&+ \sum_{{\bf k},{\bf q}}F_{\bf {\bf k},{\bf q}}
M_{{\bf k},{\bf q}}\left(\tilde{h}_{\bf k-q}^{\dag} \tilde{h}_{\bf k}
\alpha_{\bf q}^{\dag} + \mbox{H.c.}\right)\ , \nonumber
\end{eqnarray}
where $E_{\bf k}$ and $\omega_{\bf q}$ are the polaron and magnon energies,
respectively; $V_{{\bf k},{\bf k^{\prime}},{\bf q}}$ is the ``direct''
polaron-polaron interaction; $M_{{\bf k},{\bf q}}$ is the ``bare'' 
hole-magnon vertex; $F_{\bf {\bf k},{\bf q}}$ is the renormalization 
form factor.

The two-hole problem for this effective Hamiltonian has been considered 
in Ref. \cite {bel97}, and the ground state was found to be a 
bound state of $d_{x^2-y^2}$ symmetry for all $J/t > 0.2$. 
In terms of the spin polaron operators,
the wave function of the $d$-wave bound state with the total momentum 
${\rm {\bf P}}=0$ can be written as 
\begin{equation}
|\Psi_{{\rm {\bf P}}=0}^{d}\rangle=\sum_{\bf k}
\Delta^{d}_{\bf k}\tilde{h}_{\bf k}^{\dag}
\tilde{h}_{\bf -k}^{\dag} |0\rangle \ .
\label{eq:dwave}
\end{equation}
The allowed functional form of $\Delta^{d}_{\bf k}$ ensures 
the $d$-wave symmetry of the ground state. Furthermore,
due to the fact that the ground state has broken AF 
symmetry,
$\Delta^{d}_{{\bf k} + (\pi,\pi)} = -\Delta^{d}_{\bf k}$.
Consequently, this imposes that the {\em centres} 
of the polarons are always on different sublattices, which in turn 
guarantees that the total $S^z = 0$. In what follows we show that
our comparison leads us to the conclusion that the large higher harmonics
of $\Delta^{d}_{\bf k}$ play an important role in
determining the behaviour of $C(r)$ and $\langle n_{\bf q} \rangle$.

The binding energies of the $d$-wave bound state obtained numerically
by ED and analytically from the canonical transformation work, for 
the experimentally realistic $J/t=0.3$, 
are $E_b^{ED}/t\simeq-0.05$ and
$E_b^{CT}/t\simeq-0.02$, respectively. The reasonable agreement of these 
energies supports the idea that the systems being studied 
represent the same physics.

Figure \ref{fig:Cofr} shows our results for the hole-hole correlation
function $C(r)$ in the $d$-wave bound state. 
Very similar trends are found in both results: 
they both show that about
45\% of the time the holes prefer to stay at the 
nearest and next-nearest neighbour distances. That our analytical work
produces such behaviour is not inconsistent with our statements
regarding the form of $\Delta^{d}_{\bf k}$: the {\it centres} of the polarons
are indeed restricted to be on opposite sublattices, but 
the holes are almost equally distributed on both sublattices,
with the maximum probability of separation being at $\sqrt{2}$. In fact,
our analysis of the harmonics in $\Delta^{d}_{\bf k}$ shows that about 50\% 
of the polarons in the bound state are separated by one lattice constant.
Then, the peak of $C(r)$ at $r=\sqrt{2}$ arises from ``strings" of length one.
The reason for such behaviour is in the mentioned hole dressing
due to hoppings. Put another way, the quasiparticles are something
akin to a ``spin bag" with the moving hole inside the bag.  
Thus, the spin-polaron picture provides 
a natural explanation for the
$\sqrt{2}$-paradox found here and in earlier numerical studies \cite{prd94}.

Table I  shows a comparison of the ED and analytical electron
distribution functions $\langle n_{\bf q} \rangle$. As we found with
$C(r)$, both the numerical and analytical quantities show similar trends.
A full explanation of the ED $\langle n_{\bf q} \rangle$ behaviour 
based on the superposition of the effects of (i) the 
$d$-wave symmetry bound state and (ii) of the internal 
structure of the 
polarons will be presented elsewhere. Here
we focus on the features along ABZ boundary which are not
disguised by kinematic effects and can be directly 
connected with
the form of $\Delta^{d}_{\bf k}$.  
Our analytical and numerical results  have a maximum at  
${\bf k} = (\pi/2,\pi/2)$, and have minima at $(3\pi/4,\pi/4)$
and $(\pi/4,3\pi/4)$. 
The first feature may be understood as an effect arising from
nothing more than the $d$-wave symmetry of the bound state. The
electron momentum distribution function is reduced from its
half-filled value of 1/2 when holes occupy that momentum state. 
However, the bare hole number is proportional to $(\Delta^{d}_{\bf k})^2$, 
which is identically zero at $(\pi/2,\pi/2)$. Thus,
$\langle n_{\bf q} \rangle$ 
must show a maximum at this wave vector, so one
cannot find hole pockets for a $d$-wave symmetry bound state.

In future work we will report on the spin correlations that are present
in this two-hole ground state. 
Further, generalizations of the $t$-$J$ to forms that are
believed to better represent the doped ${\rm CuO_2}$ planes have been
investigated, and this work will also be discussed elsewhere.
This work was supported by the RGC of Hong Kong, and the NSERC of Canada. 

\vspace*{-0.5cm}

\begin{table}
\caption{The electron momentum distribution function, 
$\langle n_{\bf q} \rangle$, for $J/t = 0.3$, obtained from
our ED numerics and the CT analytical work.}
\begin{tabular}{r@{}lcc}
\multicolumn{2}{c}{\bf q}&
$\langle n_{\bf q} \rangle$ (ED)
&$\langle n_{\bf q} \rangle$ (CT) \\ 
\hline 
(0,&0) & 0.528 & 0.500 \\ 
$(\pi/4$,&$\pi/4)$ & 0.527  & 0.514  \\ 
$(\pi/2$,&$\pi/2)$ & 0.494  & 0.500  \\ 
$(3\pi/4$,&$3\pi/4)$ & 0.408  & 0.429  \\ 
$(\pi$,&$\pi)$ & 0.406  & 0.393  \\ 
$(\pi/2$,&$0)$ & 0.521  & 0.516  \\ 
$(\pi$,&$0)$ & 0.482  & 0.495  \\ 
$(\pi$,&$\pi/2)$ & 0.395  & 0.451  \\ 
$(3\pi/4$,&$\pi/4)$ & 0.466  & 0.477  \\
\end{tabular}
\end{table}

\end{document}